\documentclass[10pt]{iopart}

\PassOptionsToPackage{linktocpage}{hyperref}
\usepackage{iopams} 
\usepackage{graphics}
\usepackage{graphicx} 
\usepackage{graphicx}
\usepackage{dcolumn}
\usepackage{bm}
\usepackage{color}
\usepackage{latexsym}
\usepackage{float}
\usepackage[T1]{fontenc}

\input{epsf}
\begin{document}

\title{SANS Study of Vortex Lattice Structural Transition in Optimally Doped (Ba$_{1-x}$K$_{x}$)Fe$_{2}$As$_{2}$}

\author{S. Demirdi\c{s}}
\address{J\"ulich  Center for Neutron Science (JCNS) at Heinz Maier-Leibnitz Zentrum (MLZ), Forschungszentrum J\"ulich GmbH, Lichtenbergstrasse 1, D-85747 Garching, Germany} 
\author{C.J. van der Beek}
\address{Laboratoire des Solides Irradi\'{e}s, CNRS UMR 7642 \& CEA-DSM-IRAMIS, Ecole Polytechnique, F91128 Palaiseau cedex, France}
\author{S. M\"uhlbauer}
\address{Technische Universit\"at M\"unchen, Forschungsneutronenquelle Heinz Maier-Leibnitz (FRM II) \\D-85748, Garching, Germany}
\author{Y. Su}
\address{ J\"ulich  Center for Neutron Science (JCNS) at Heinz Maier-Leibnitz Zentrum (MLZ), Forschungszentrum J\"ulich GmbH, Lichtenbergstrasse 1,D-85747 Garching, Germany} 
\author{Th. Wolf}
\address{Karlsruher Institut f\"ur Technologie, Institut f\"ur Festk\"orperphysik, D-7602, Karlsruhe, Germany}

\ead{s.demirdis@fz-juelich.de}
\vspace{10pt}
\begin{indented}
\item[]May 2016
\end{indented}

\begin{abstract}
Small-angle neutron scattering on high quality single crystalline Ba$_{1-x}$K$_{x}$Fe$_{2}$As$_{2}$ reveals the transition from a low--field vortex solid phase with orientational order to a vortex polycrystal at high magnetic field. The vortex order-disorder transition is correlated with the second-peak feature in isothermal hysteresis loops, and is interpreted in terms of the generation of supplementary vortex solid dislocations. The sharp drop of the structure factor above the second peak field is explained by the dynamics of freezing of the vortex ensemble in the high field phase.
\end{abstract}

%
%
%
%
\ioptwocol

The relation between vortex lattice structure, vortex dynamics, and the vortex phase diagram has been the object of intensive studies in intermetallic \cite{Paltiel2000,Menghini2002,Fasano2002,Klein2009}, cuprate \cite{Cubitt93,Zeldov95,vdBeek2000}, and, recently, Fe-based type II superconductors \cite{vdBeek2010}. This has resulted in the paradigm in which, on the one hand, for weakly disordered type II superconductors, a long-range ordered low temperature vortex solid state transits via a first order melting transition, to a high-temperature vortex liquid lacking long-range superconducting  order \cite{Zeldov95}. On the other hand, disordered materials with strong vortex pinning  exhibit a continuous ``glass'' transition from a low-temperature disordered vortex state to the vortex liquid \cite{Klein2009}. The nature (and, indeed, the existence)  of the ``vortex glass transition'' is not definitively settled \cite{Espinosa-Arronte}. At best, gauge symmetry is broken as the vortex glass forms, while the melting transition breaks  gauge symmetry as well as translational and orientational order. The case of intermediate disorder is interesting, since three vortex phases potentially appear. Upon heating, the ordered vortex solid transits to the vortex liquid. However, an increase in vortex density at low temperature is also followed by a first order transition, to a second vortex solid (or glassy) phase with strongly modified dynamics \cite{Klein2009,Chikumoto92,Kokkaliaris,Deligiannis,vdBeek2000}.  The change in current-voltage characteristics that this entails leads to the so-called ``second magnetization peak'' (SMP)  phenomenon in numerous type II superconductors. The first order transition at the SMP onset field, $B_{sp}$ \cite{Deligiannis,Kokkaliaris}, has been interpreted either in terms of a structural change from a dislocation-free vortex ``Bragg glass'' to a plastically disordered phase \cite{Giamarchi,Giamarchi2,Klein}, or in terms of the loss of vortex integrity along the field direction \cite{vdBeek2000,Colson2003}. Thus, the understanding of the vortex phase diagram in disordered superconductors is incomplete at best, unsettled questions being the mechanism of the SMP transition, the link between vortex solid structure and dynamics, the nature of the vortex glass, and whether the high-field vortex solid is, in all cases, distinct from the liquid.

The SMP has been observed earlier in many other systems, and has been interpreted either as a structural transition from an ordered to a disordered state, or as a loss of vortex correlation along the field direction. Depending on the system (weakly disordered/or strong pinning) this structural transition drives the vortex lattice to a liquid phase with a melting transition or to a disordered solid phase with a glass transition respectively \cite{Ling,Kes,Klein,Paltiel2000,Cubitt93,Kokkaliaris}. 
In conventional superconductors like in Nb single crystal  and  in NbSe$_2$ an anomalous phenomena is found in the vicinity of the peak effect where the critical current $j_c$ increases sharply below the upper critical field H$_{c2}$ \cite{Gammel,Kokkaliaris}. The vortex lattice  studied with SANS and $in~situ$ magnetic susceptibility measurements in high purity Nb single crystal has revealed the history dependence of the structure in the peak effect regime via ZFC and FC procedures applied  during the SANS measurements. Metastable phases of vortex matter, supercooled vortex liquid and superheated vortex solid, have been identified. The results have been interpreted as a direct structural evidence for a first-order vortex solid-liquid transition at the peak effect \cite{Ling}. However in 2002 Forgan {\em et.al} reported that in pure Nb the flux lattice structure is stable against thermal fluctuations over essentially all of the mixed state region, and  that their SANS measurements confirm the Abrikosov
picture near H$_{c2}$. They explained that the discrepancy may possibly arise from differences in sample purity and pinning \cite{Forgan}. In the high-$T_c$ superconductor Bi$_2$Sr$_2$CaCu$_2$O$_{8+x}$ the vortex lattice has been studied via SANS \cite{Cubitt}  with some other complementary technique such as Muon-spin rotation and Hall-sensor arrays magnetometry by P. H. Kes and colleagues \cite{Kes}.  They reported that Muon-spin rotation and  SANS experiments gave the first indications for an abrupt change of the vortex lattice structure along a line  in the ($B$-$T$) phase diagram. While Hall-sensor arrays  experiments proved that the flux line lattice experiences a first order phase transition (FOT) at this line. The nature of the crossover at high fields and the underlying mechanism was unresolved. However the role of the disorder was reported to be crucial and  the crossover to be a disorder driven true thermodynamic phase transition which they described as the  proliferation of dislocations in the pinned vortex lattice (low field) transforming it into an amorphous vortex glass of individually pinned pancake vortices (high field). The correlation between the observed decrease of $F(q,T)$ and the preparation procedure of the vortex ensemble, i.e. the path taken through the $(B,T)$ phase diagram before the diffraction signal is acquired is also a  relevant point. This has been little studied, and its importance is one of the main points made by our manuscript. We note that the manner of preparing the vortex ensemble is largely irrelevant in Bi$_2$Sr$_2$CaCu$_2$O$_{8+x}$, in which FC takes place through a vortex liquid state with vanishing critical current, and the vortex ensemble only freezes at the melting line.  In Ba$_{1-x}$K$_x$BiO$_3$ the preparation procedure of the vortex lattice has been studied in some detail, and it was found that, if performing temporal oscillations of the magnitude of the magnetic field during FC improves the quality of the vortex lattice, it does not change the conclusions of that particular work that the vanishing of the neutron diffraction form factor coincides with the SMP onset \cite{Klein}.
\begin{figure}
\includegraphics[width=0.5\textwidth]{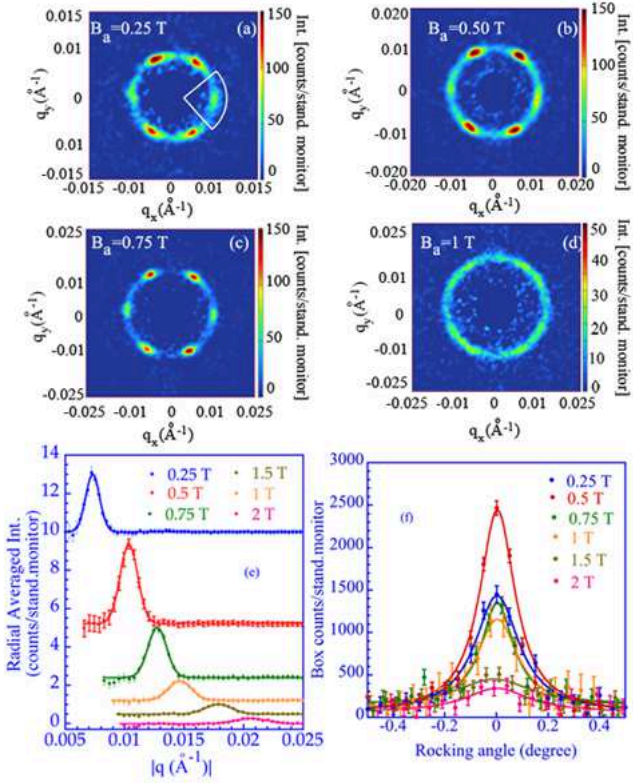}
 \caption{ (a-d) Small-angle neutron scattering pattern of the vortex solid in Ba$_{0.64}$K$_{0.36}$Fe$_{2}$As$_{2}$, for magnetic fields ranging from 0.25~T to 1~T. The patterns represent sums over a rocking scan with respect to the vertical axis, with a zero field background at 45 K subtracted. The weaker intensity of the spots on the horizontal line is due to the larger distance from the rocking axis (Lorentz factor).  The direct beam is masked for better visibility; the data have been smoothed using a Gaussian filter. (e) Radial intensity distribution versus  $|q|$, for different applied magnetic fields. The baselines of each curve have been offset for visibility. (f) Angular dependence of the diffracted intensity (rocking curves) at different applied magnetic fields. The solid lines are fits to the data  with a  Lorentzian function.  }
\label{fig:SANS}
 \end{figure}

In the more particular case of Fe-based superconductors several  techniques have revealed the strong pinning properties and the related highly disordered vortex ensembles. Apart from  the work on KFe$_2$As$_2$ \cite{furukawa}, and on  BaFe$_2$(As$_{0.67}$P$_{0.33}$)$_2$ \cite{morisaki}  where a mosaic of single crystals have been annealed right before the experiment, real space imaging and Small Angle Neutron Scattering  (SANS) studies of single crystalline Fe-based superconductors have consistently revealed highly disordered vortex ensembles over a very wide range of magnetic fields  \cite{Eskildsen,demirdis,demirdis2}, the single crystalline (Ba$_{1-x}$K$_{x}$)Fe$_{2}$As$_{2}$ material investigated below being no exception. Different studies at very low fields mention either short-range triangular order, or highly disordered vortex configurations characterized by local triangular clusters and vortex chains \cite{Shan,Yang}.

 \begin{figure}[h]
\includegraphics[width=0.5\textwidth]{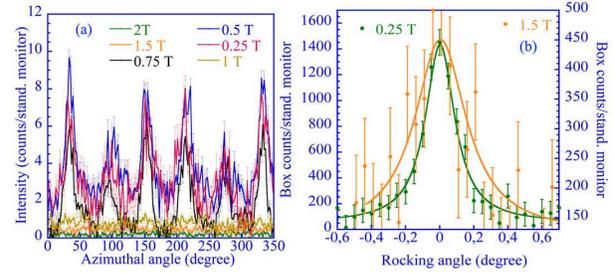}
\caption{(Color online)  (a) Azimuthal intensity distribution as a function of increasing field (b)  Angular dependence of the diffracted intensity   for applied fields 0.25 T and 1.5 T showing the broadening in the FWHM of the rocking curves . The solid lines are fits to the data  with a  Lorentzian function.}
\label{fig:Azimuthal-rock}
\end{figure}

Nevertheless, a SMP feature has been observed  \cite{Wang,Shen,Sugui}.  The SANS results on Ba$_{0.64}$K$_{0.36}$Fe$_{2}$As$_{2}$ single crystals presented below reveal that vortex lattice disorder can be sufficiently weak to permit the observation of  well-defined Bragg peaks corresponding to a long-range orientationally ordered triangular lattice. However, the Bragg peak shape betrays considerable remaining disorder. A confrontation with the vortex phase diagram established using vibrating sample- and SQUID magnetometry reveals that the decrease with field of the vortex structure factor is governed by the manner in which vortices are frozen in the {\em high--field state}. The latter turns out to be a vortex polycrystal, such as in Refs.~\cite{Klein2009,Troianovski}.

SANS experiments were carried out on the SANS-1 \cite{sans} instrument co-operated by Technische Universit\"at M\"unchen (TUM) and Helmholtz-Zentrum Geesthacht (HZG) at the research reactor of the Heinz Maier-Leibnitz Zentrum  in Garching, Germany. Neutrons with wavelengths  $\lambda_n= 5.5$ to $8$ {\AA } were used, and the wavelength spread was  $\Delta\lambda_{n}/\lambda_{n}=0.1$.  We used a  large ($10\times10\times2$ mm$^3$) Ba$_{0.64}$K$_{0.36}$Fe$_{2}$As$_{2}$ single crystal with critical temperature $T_c$= 38 K, grown using a self-flux method \cite{crystal},  and that has not undergone any post-growth treatment. The sample was mounted in an Al container sealed under He atmosphere in order to avoid any exposure to air. The vortex lattice was obtained by applying the desired magnetic field (between 0.25 T-2 T) above $T_{c}$ and subsequently cooling to 4~K. For each experimental configuration, the zero-field background was measured at $T=45$~K, and subtracted from the field-cooled (FC)  data. The data are evaluated using GRASP package \cite{Grasp}.  Magnetization measurements were performed on a smaller single crystal (of dimensions $2\times 3 \times 0.2$~mm$^{3}$) from the same batch using the vibrating sample magnetometry (VSM) option of  a Quantum Design Physical Property Measurement System (PPMS) and a SQUID magnetometer. These two measurement techniques differ mainly in the time scale on which the applied field is ramped:  $dB_{a}/dt = 0.6$ T/min for the VSM, against 0.03 T/min for the SQUID magnetometer.

\begin{figure}[h]
\includegraphics[width=0.5\textwidth]{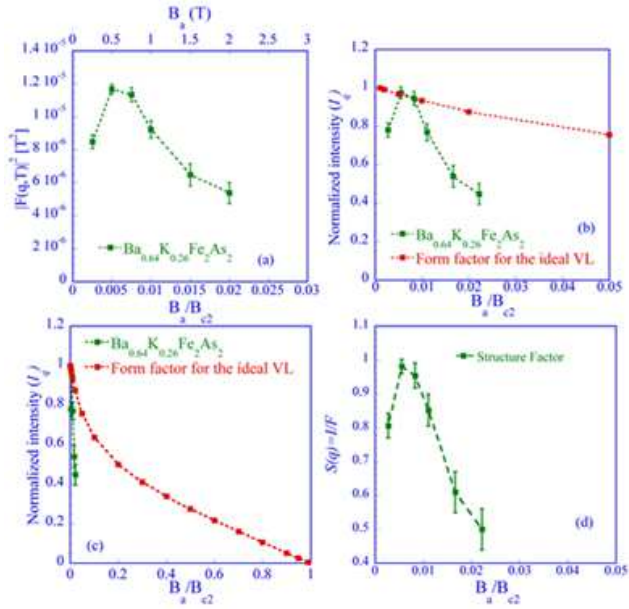}
\caption{(Color online)  (a)  Field dependence of $\mid F(q,T)\mid ^{2}$ extracted from the integrated intensity obtained by integration over an arc of 60$^\circ$ spanning the vortex Bragg peak as indicated in Fig.\ref{fig:SANS}a. (b) Data normalized by the prediction $I_{q}$ for  a perfect vortex lattice  \cite{Brandt}). (c) As (b), but over the full magnetic field range up to $B_{c2}$. (d) Field dependence of the vortex structure factor, extracted from the data in (b) using Eq.~\ref{structurefactor}. }
\label{fig:formfactor}
\end{figure}
Figure~\ref{fig:SANS} a-d shows representative diffraction patterns, measured in selected applied magnetic fields $B_{a}$ between 0.25 and 1~T, applied parallel to the $c$-axis of the single crystal, and nearly parallel to the neutron beam. The images represent sums over the rocking scans with respect to the vertical axis. The weaker intensity of spots on the horizontal line is due to the larger distance from the rocking axis (Lorentz factor). The direct beam is masked and the data is smoothed with a $2\times2$ pixel Gaussian. 
Clear diffraction peaks are observed up to $B_{a} = 0.75$~T. Above this field, the diffraction spots start to broaden, and their intensity diminishes.  A near-circular (polycrystalline) diffraction pattern is observed at $B_{a} = 1$~T, whence the scattered intensity has all but vanished for $B_{a} =2$~T.  The radial scans of the scattered intensity as function of the magnitude of the $(\frac{1}{2},\frac{1}{2}\sqrt{3})|q|$ vector (Fig.~\ref{fig:SANS}e) quantify this decrease. Figure~\ref{fig:SANS}f presents the rocking curves for different $B_{a}$ up to 2 T. The solid lines represent fits to the Lorentzian function. Here note that for the two lowest field the rocking curves can be fitted both with a Gaussian and a Lorentzian function and the fit outputs stay the same.

Figure~\ref{fig:Azimuthal-rock} a presents the Azimuthal intensity distribution as a function of increasing field showing the amplitude of the diffuse scattering intensity in between well define Bragg peaks. Figure~\ref{fig:Azimuthal-rock} b presents a comparison of the FWHM of the rocking curves for applied fields 0.25 T and 1.5 T showing the clear broadening of the rocking width with increasing field.

The average vortex lattice structure factor $S$ can be obtained from the integrated intensity of the vortex lattice Bragg peaks, 
\begin{equation}
I=F^{2} S = F^{2}(T)  \int dq_{x} \int dq_{y} S(q_{x},q_{y},K_{0}\omega),
\label{structurefactor}
\end{equation}
by correction for the vortex form factor $F$, that reflects the magnetic field distribution around a vortex  \cite{Klein}. Here, $K_0$ is the vortex reciprocal lattice vector and $\omega$ is the rocking angle. Since the sharp Bragg peaks smoothly cross over to a diffraction ring, we choose to determine the integrated intensity $I$ (integration of the rocking curve) by averaging over an arc of 60$^{\circ}$ spanning a single Bragg peak (as indicated on Fig.~\ref{fig:SANS}a). This is then corrected for the detector efficiency, the number of monitor counts, and the sample transmission, to yield the vortex Bragg peak intensity 
\begin{equation}
I_{q}=2\pi V \phi \left (\frac{\gamma}{4}\right )^{2}\frac{\lambda^{2}_{n}}{\Phi^{2}_{0}|q|}\mid F(q,T)\mid ^{2}.
\label{formfactor}
\end{equation}
Here $V$ is the illuminated sample volume, $\phi$ the neutron flux density,  $\gamma =1.91$~$\mu_{N}$ the neutron magnetic moment, $q$ a vector in reciprocal space. $F(q,T)$ depends on temperature through the penetration depth $\lambda_{L}(T)$ and the coherence length $\xi(T)$, and decreases with $B_{a}$ because of the gradual weakening of the internal field modulation due to vortex overlap. The Ginzburg-Landau coefficient is estimated for this compound $\kappa=\lambda/\xi\approx100$  based on data from Refs. \cite{Putti,Martin}. Figure~\ref{fig:formfactor}a shows $F^2(q,T)$ versus the reduced magnetic field $B_{a}/B_{c2}$. The value of the upper critical field $B_{c2}$  has been taken from Refs.\cite{Wang},\cite{Altarawneh}. The measured intensity has its maximum value at low fields,  and drops abruptly above 0.5 T, well below $B_{c2}$.  Figs.~\ref{fig:formfactor} b-c show a comparison of the present data with a numerical solution of the Ginzburg Landau equations from $B_{c1}$ up to $B_{c2}$ for the case of the triangular perfect lattice \cite{Brandt}. The latter gives an analytical extension to calculate the Frourier coefficients of the triangular  vortex lattice for any induction $10^{-3}<B_{a}/B_{c2}<1$ and to all relevant GL parameters. Data  presented with red markers giving the form factor for an ideal VL in  Fig.~\ref{fig:formfactor} b-c has been taken from the calculated values of first five Fourier coefficients $b_k=b_{mn}$ of the triangular VL for the $\kappa\gg1$ limit divided by the London limit
$b_k$= $\bar{B}$/1+$({K^2}\lambda^2)$, as described in Ref.\cite{Brandt}. Since, for the ideal VL case the structure factor
 $S(q)=1$, equation(\ref{structurefactor}) will reduce to $I_q=F^2$. A normalization by $I_q$ for the perfect lattice  allows one to extract the averaged structure factor $S$ of the vortex ensemble, see Figs.~\ref{fig:formfactor} b-d. The sharp decrease of $S$ above $B_{a} \approx 0.5$~T  indicates a structural disordering transition of the vortex solid.

\begin{figure}[t]
\includegraphics[width=0.5\textwidth]{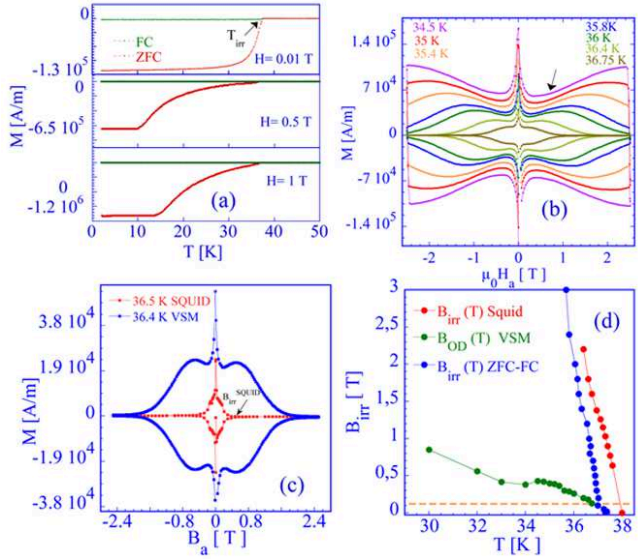}
\caption{(Color online) (a) representative curve of the dc magnetization measured under ZFC and FC conditions, for various fields. The arrow indicates the irreversibility temperature $T_{irr}$.  (b)  Loops of the hysteric magnetization $M(B_{a})$ of the Ba$_{0.64}$K$_{0.36}$Fe$_{2}$As$_{2}$ single crystal, for selected different  temperatures. The onset of the SMP is indicated by the arrow. (c) Hysteresis loops measured at $T= 36.5$~K using the VSM and SQUID techniques. (d) Magnetic phase diagram of the vortex ensemble in the (Ba$_{0.64}$K$_{0.36}$)Fe$_{2}$As$_{2}$ single crystal. Here the  vertical yellow line indicates  the low-field phase (below the orange line)  and the high-field phase (above the orange line) of the vortex ensemble. The order-disorder field $B_{OD}$ has been defined as the onset of the SMP. The difference in irreversibility fields extracted from isothermal SQUID and ZFC-FC measurements underscore the effect of flux creep.}
\label{fig:diagram}
\end{figure}

In order to correlate this transition with the vortex phase diagram, we resort to magnetization measurements. Figure~\ref{fig:diagram}a shows representative curves of  the temperature-dependent dc magnetization for  Ba$_{0.64}$K$_{0.36}$Fe$_{2}$As$_{2}$, measured between $T= 4.2$ K and 50~K and magnetic fields  from 0.01 to 3~T, for zero-field cooling (ZFC) and successive field cooling (FC, after warming) protocols. The demise of vortex pinning by material defects is signaled by the merger of the ZFC and FC curves at the irreversibility temperature $T_{irr}(B)$. Selected curves showing the temperature evolution of  the isothermal hysteretic loops of the dc magnetization measured using the VSM are presented in Fig.~\ref{fig:diagram}b. As usual, the magnetic hysteresis can be interpreted in terms of the Bean critical state model, in which the vortex distribution inside the superconducting crystal is macroscopically  inhomogeneous and history--dependent due to pinning \cite{vdBeek2010}. At all temperatures, the irreversible magnetization of  Ba$_{0.64}$K$_{0.36}$Fe$_{2}$As$_{2}$ features the zero-field peak (central peak) found in all  Fe-based superconductors \cite{Yang,Shen,vdBeek2010,Pramanik}, which we interpret  in terms of strong vortex pinning by sparse nanometric defects \cite{vdBeek2010,demirdis}. 
Below 30 K, strong pinning dominates, and no SMP is observable. At higher temperatures, the central peak gives way to collective pinning by atomic-sized point defects \cite{vdBeek2010} and to the increase of the magnetic moment (SMP) at the onset field $B_{OD}$ \cite{Deligiannis,Kokkaliaris,Klein2009,vdBeek2010}, the temperature dependence of which is shown in Fig.~\ref{fig:diagram}d.  

Fig.~\ref{fig:diagram}c compares magnetic hysteresis loops measured at nearly the same temperature using the (fast) VSM vs. the (slow) SQUID apparatus. In the first case, the hysteretic part of the magnetic moment ({\em i.e.} pinning) only vanishes far above $B_{OD}$. In the latter case, the effects of pinning vanish at $B_{irr}^{SQUID} \sim 0.4$~T, with the SMP unobservable.  This confirms that the effects of thermally activated flux creep are very pronounced in this material \cite{marcin} -- in the case of the SQUID experiment,  sufficiently so to {\em  bring the vortex ensemble to thermodynamic equilibrium} above 0.4~T.  In contrast to $B_{irr}$, the SMP onset ($B_{OD}$) is not affected by flux creep (see Fig.~\ref{fig:diagram}d), suggesting that it corresponds to an intrinsic ``order-disorder'' transition of the vortex ensemble  Refs.~\cite{Paltiel2000,Deligiannis,Kokkaliaris,Paltiel2,Klein2009,vdBeek2000}.

To interpret the SANS patterns, we stress that these are obtained using a FC protocol and therefore {\em reflect the vortex ensemble as quenched at}  $T_{irr}(B)$, which plays the role of a ``freezing temperature'' $T_f$  \cite{demirdis}.  From the temperature dependence  of the diffracted intensity at 0.25 T, we could conclude that the FWHM of the obtained rocking curves remain constant  below 32.5 K upon cooling at low temperatures. This gives an experimental evidence for $T_f$ , the temperature value at which the vortex lattice freezes and does not move upon further cooling during the FC procedure of the SANS experiment.

For the experiments performed at  $B_{a} \lesssim 0.25$~T,  $T_{irr}(B) \gtrsim T_f$ coincides with the  $B_{OD}(T)$ boundary (below the orange line in  Fig.~\ref{fig:diagram}d). Field cooling across the high-field vortex state has no effect: thermal activation equilibrates the vortex positions on the time scale of the experiment, and only at $B_{OD}(T)$ the vortex positions are fixed. The diffraction patterns therefore represent the (ordered) vortex ensemble (with $S \simeq 1$) such as  this is quenched directly into the low--field state. At intermediate fields, $0.25$~T~$\lesssim B_{a} \lesssim 1$~T, cooling through the high field state does affect the vortex positions to a greater or lesser extent. This results in the progressive disordering of the vortex ensemble before this is quenched into the low-field state, and the concomitant decrease of the structure factor with increasing $B_{a}$. Above 1~T the applied field exceeds $B_{OD}(T)$ at all $T$. The low field state is never reached, and the vortex ensemble is quenched as a disordered polycrystal before being definitively fixed by strong pinning below $T \lesssim \frac{1}{2} T_{c}$.  Therefore, the observed decrease of the structure factor does not as much reflect the structural properties of the low--field vortex state, as the dynamics of vortex freezing in the high field state.

\begin{figure}[t]
\centering\includegraphics[width=0.5\textwidth]{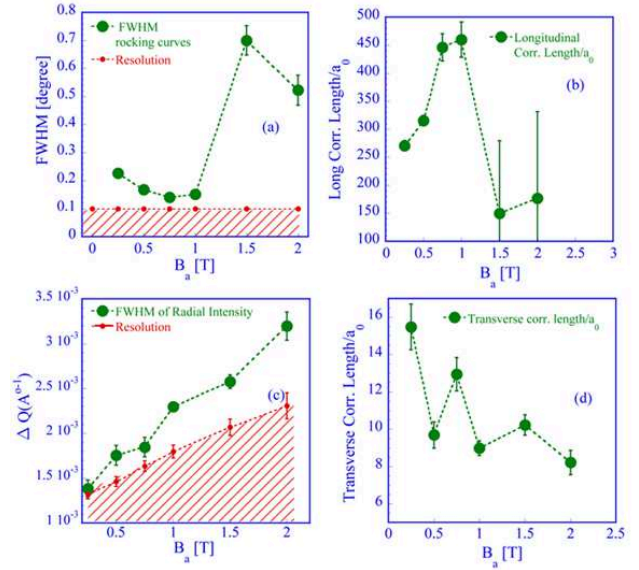}
\caption{(Color online) (a) Full width at half-maximum of the rocking curves extracted from the fits in Fig.\protect\ref{fig:SANS}f. (b) Field dependence of  the longitudinal correlation length extracted using the FWHM of rocking curves. Here the resolution is given by the beam divergence only. (c) Full width at half-maximum of the radially averaged intensity, extracted from the fits in Fig.\protect\ref{fig:SANS}e. (d) Field dependence of the transverse correlation length  extracted from the width of the radial averaged diffracted intensity. Here red shaded areas represent  the resolution limit of the SANS instrument. }
\label{fig:corrlength}
\end{figure}

We now turn to the quantitative analysis of the vortex structure. Figure~\ref{fig:corrlength}a shows the full width at half maximum of the rocking curves extracted from the fits in Fig.~\protect\ref{fig:SANS}f, compared to the resolution limit of the SANS setup. The instrumental resolution of rocking scans (longitudinal direction parallel to the magnetic field) is - at a given Bragg angle - well described by the divergency of the incident neutron beam of $\sigma_{res \parallel}=0.11^{\circ}$. The width of the rocking curves varies slightly up to 1~T before increasing sharply for higher fields. From the rocking curve width corrected for the experimental resolution, $\sigma^{2}_{m}=(\sigma^{2}_{rock}-\sigma^{2}_{res \parallel})$ we extract the longitudinal correlation length $\xi_{\parallel}=2 \pi/q\sin({\sigma_{m}})$ \cite{Yaron,Grigoriev}. $\xi_{\parallel}$, shown in Fig.~\ref{fig:corrlength}b,  measures the distance over which the vortex relative displacements along the field direction remain smaller than the vortex spacing $a_{0}$.  For low fields up to $B_a = 1$~T, $\xi_{\parallel}$ has a value of several hundred $a_0$, decreasing significantly above this field value.

The radial width of the Bragg peaks is limited by the resolution $\sigma_{res \perp}=\sqrt{4\pi^{2}( \delta\theta/\lambda_{n})^{2}+q^{2}(\Delta\lambda_{n}/\lambda_{n})^{2}}$ with the divergency of the beam in radian $\delta\theta=\pi\sigma_{res \parallel}/180^{\circ}$. From the radial widths of the Bragg peaks as a function of $B_{a}$, plotted in Fig.~\ref{fig:corrlength}c, we determine the transverse correlation length $\xi_{\perp}=2\pi/(\sigma^{2}_{q}-\sigma^{2}_{res \perp})^{1/2}$  \cite{Eskildsen}. This measures the distance beyond which vortex relative displacements perpendicular to the field direction exceed $a_{0}$ (Fig.~\ref{fig:corrlength}d). 

  \begin{figure}[h]
\centering\includegraphics[width=0.3\textwidth]{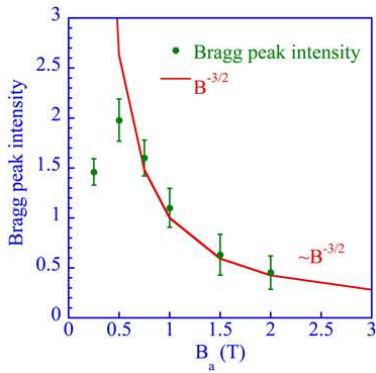}
\caption{(Color online) Normalized Bragg peak intensity that decreases as $\sim$$B^{-3/2}_{a}$.  }
\label{fig:peakintensity}
\end{figure}

The large values of the correlation lengths are compatible with what would be expected from weak collective pinning in the so-called ``random manifold regime'' \cite{Giamarchi,Giamarchi2}, but not with strong pinning. The SANS results are thus representative of the vortex structure such as this is frozen at high $T$. The high field results thus reveal that the vortex state above $B_{OD}(T)$ is a vortex polycrystal. 

As for the nature of the low--field state, we compare the shape and field--dependence of the observed vortex Bragg peaks to the Bragg--glass predictions \cite{Giamarchi,Klein}. These hold that (i) the Bragg peak intensity has a power law tail, $I \sim q^{3-\eta}$, with $\eta \approx 1$, (ii) the Bragg peak height decreases proportionally to the transverse correlation length $\xi_{\perp}$, while the full width at half maximum stays constant, and  (iii) the Bragg peak intensity decreases as $B_{a}^{-3/2}$ \cite{Giamarchi,Klein}. 
In our experiment, only the third prediction is verified by the data (see figure~\ref{fig:peakintensity}). However, given that the decrease of the Bragg peak height and the structure factor are determined by the properties of the high--field, and not of the low--field state, our present data is insufficient to confirm whether a Bragg glass is formed below $B_{OD}$ or not.
In summary, we have shown that in optimally doped single crystalline Ba$_{1-x}$K$_{x}$Fe$_{2}$As$_{2}$, pinning disorder at high temperature may be sufficiently weak to permit the formation of a low--field vortex ensemble with long-range orientational order. Quantitative analysis does not confirm the presence of a Bragg glass. Rather, the observed decrease of the vortex structure factor is the result of the dynamics of vortex freezing through the high--field state. The low--field vortex state gives way, by means of structural transition at the second peak onset $B_{OD}$, to a high--field polycrystal characterized by the lack of orientational order and a much reduced correlation length parallel to the magnetic field. The order-disorder transition of the vortex ensemble is necessarily mediated by the generation of supplementary vortex lattice dislocations, and is therefore very similar to that previously observed in NbSe$_{2}$ \cite{Paltiel2000,Menghini2002,Fasano2002,Troianovski} and MgB$_{2}$ \cite{Klein2009}. 

\section*{Acknowledgements}
This work is based upon experiments performed at SANS-1 \cite{SANS-1} instrument co-operated by Technische Universit\"at M\"unchen  (TUM) and Helmholtz-Zentrum Geesthacht (HZG) at the research reactor of  the Heinz Maier-Leibnitz Zentrum (MLZ), Garching, Germany.

\end{document}